# Thickness Effect on Fluctuation of Electron States in Thin Film and Implication to Lattice Constant Change Due to Size Reduction

Mikrajuddin Abdullah

[1)]Department of Physics, Bandung Institute of Technology

[2)]Research Center for Nanotechnology,

Bandung Institute of Technology

Jl. Ganesa 10 Bandung 40132, Indonesia

Email: mikrajuddin@gmail.com

**Abstract**

We propose a model for predicting the fluctuations of electron states in thin films as function of film thickness. The model was derived based on the assumption of the existence of potential barrier fluctuations on the film surface. Since the wave functions of electrons in the film is determined by the boundary conditions of potential on the film surface, potential fluctuations on the film surface implies the fluctuations of electron states in the film. The model was then extended to predict the effect of size on the lattice constant of thin films or nanoparticles. The derived equations can explain fairly several experimental data.

## 1. Introduction

Eigen states of the electrons in thin films are generally solved with the assumption that the electrons are in a one-dimensional potential well [1]. The potential in the well is assumed to be zero and the potential outside is assumed to be very large constant. Further simplification that commonly performed is by considering the potential wall has infinite height. With these assumptions, the boundary condition is the wave functions are zero at the potential wall and the resulting solutions are sinusoidal functions [1].

Assumption that the potential wall has an infinity height is merely idealization that commonly used to introduce the elementary quantum theory for students [1]. The height of the potential wall is actually finite. With the finite height of the walls, the electrons can still jump out of the film surface and gave rise to a number of phenomena such as the photoelectric effect, thermal emission, electric field induced electron emission and so on.

Since the film surface is always in contact with another material such as air then there is an interaction between the atoms at the film surface with the atoms of the surrounding material. Such interaction is expected to cause a fluctuation of atomic states on the film surface, especially when the states of atoms outside the film also fluctuate (for example atoms or molecules of air), which in turn causing fluctuations in the height of the potential wall. Because the eigen states of the electrons inside the potential well depend on the boundary condition of potential on the wall, fluctuations in wall potential causing fluctuations in electron eigen functions in the well. If the film is very thick or the material is in bulk state, the effect of fluctuations in the surface potential is negligible so that the fluctuations in potential well do not affect the eigen states of electrons in the well. Conversely, if the film is very thin, the surface effects become dominant such that fluctuations in potential wall affect the eigen state of electrons in the well.

According to our knowledge there has been no discussion in detail on the effect of film thickness on the eigen states of the electrons in the film. The purpose of this paper is to formulate a model for predicting the effect of film thickness on the fluctuation of eigen state of electrons in the film. Then we examine the implications for predicting the variation of lattice constant if the film thickness change. Observations of changes in the lattice constants of the materials when reducing the size is not a new phenomenon. There have been many reports about such observations as in platinum [2,3], gold [2], ceria [4], tantalum [5], titania [6]. Using a scanning high energy electron diffraction, Soliard and Flüeli observed that small particles of gold and platinum (size of 39-500 angsrom) experienced a lattice constant reduction if the particle size decreases [2]. Such phenomenon was observed at all measurement temperature from about 100 K to 600 K.

## 2. Modeling

### 2.1 Fluctuations of Electron States

We model the thin film as a one-dimensional potential well with a width L (Figure 1). The potential in the well is taken to be zero and potential outside the well is V and finite. The potential outside the well is the surface potential of the film and assumed to be slightly fluctuates.

We determine the solution of Schrodinger equation by firstly ignoring the wall potential fluctuations. We consider the wall potential height to be constant so that the eigen solution can be found easily. Having obtaining the solution, we consider the effect of wall potential fluctuations.

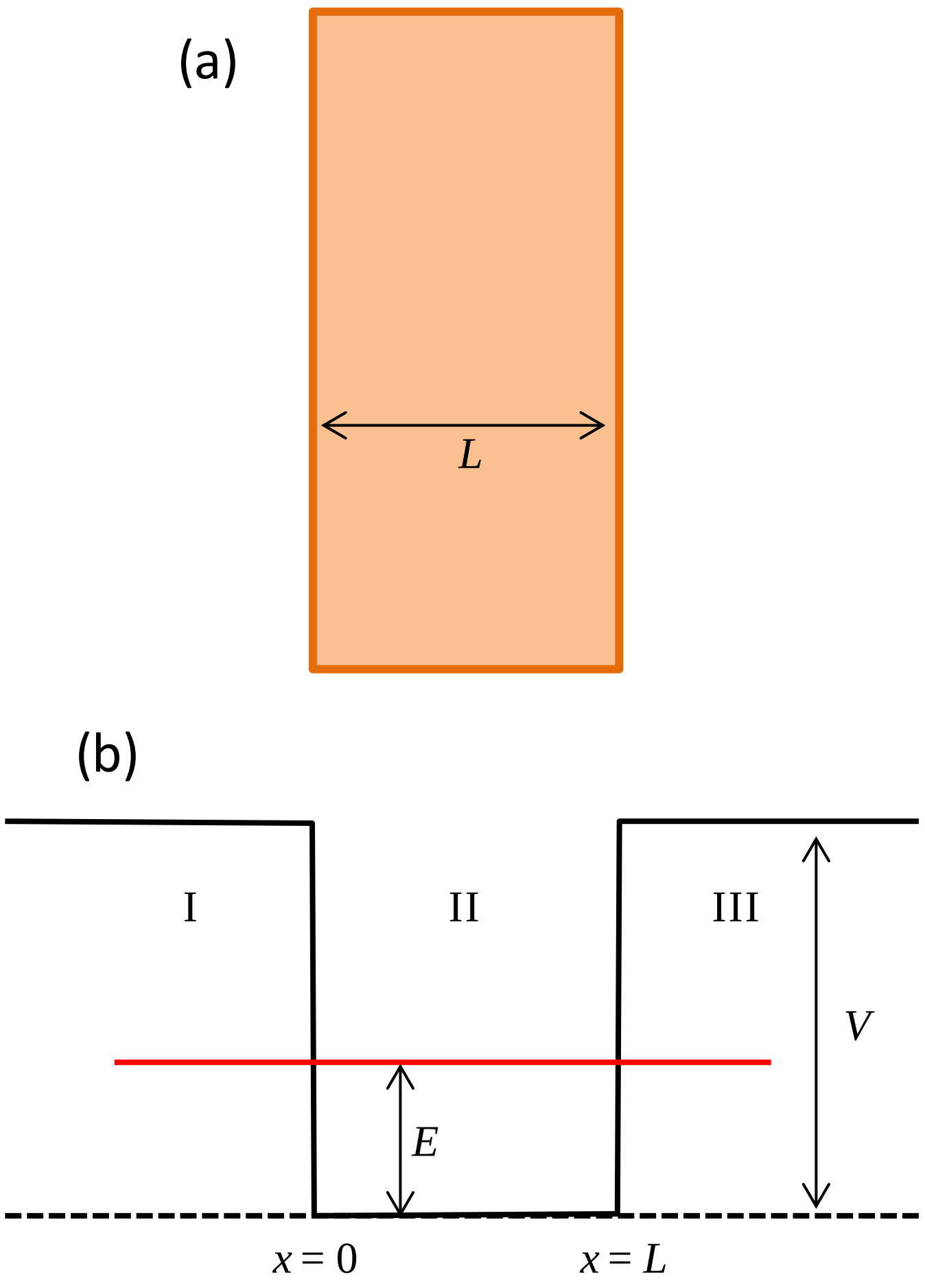

**Figure 1**. (a) Thin films and (b) models of thin film as a one-dimensional potential well with a width L. The potential in the well is 0 and the outside of the well is V. The potential is divided into regions I, II, and III. In regions I and III, E <V whereas in region II, E> V.

The Schrodinger equations for electrons in regions I, II, and III (illustrated in Figure 1) are $\frac{d^2\psi_I}{dx^2} = \alpha^2\psi_I$, $\frac{d^2\psi_{II}}{dx^2} = -\beta^2\psi_{II}$, $\frac{d^2\psi_{III}}{dx^2} = \alpha^2\psi_{III}$ with

$$\alpha = \sqrt{\frac{2m(V-E)}{\hbar^2}} \tag{1}$$

$$\beta = \sqrt{\frac{2mE}{\hbar^2}} \tag{2}$$

Since the fluctuation in the wall potential has not been taken into consideration, then α is a constant. The general solutions to the wave functions by considering that the wave function must be finite in regions I, II, and III are

$$\psi_I(x) = Ae^{\alpha x} \tag{3a}$$

$$\psi_{II}(x) = Ce^{i\beta x} + De^{-i\beta x} \tag{3b}$$

$$\psi_{III}(x) = Fe^{-\alpha x} \tag{3c}$$

with A, C, D, and F are constants which in general are complex numbers.

The boundary conditions that must be met are $\psi_I(0) = \psi_{II}(0)$, $\psi_{II}(L) = \psi_{III}(L)$, $\psi'_I(0) = \psi'_{II}(0)$, and $\psi'_{II}(L) = \psi'_{III}(L)$. Such boundary conditions yield the following equation set

$$A = B + C \tag{4a}$$

$$Ce^{i\beta L} + De^{-i\beta L} = Fe^{-\alpha L} \tag{4b}$$

$$\alpha A = i\beta(C - D) \tag{4c}$$

$$i\beta\left(Ce^{i\beta L} - De^{-i\beta L}\right) = -\alpha Fe^{-\alpha L} \tag{4d}$$

From equations (4a) - (4d) we get the following relationship

$$\left(\frac{\beta - i\alpha}{\beta + i\alpha}\right)^2 = e^{-i2\beta L} \tag{5}$$

Let us write $\beta - i\alpha = \sqrt{\alpha^2 + \beta^2}\, e^{-i\phi}$ where $\tan\phi = \alpha/\beta$. By this definition, the equation (7) can be written as $\exp[-i4\phi] = \exp[-2\beta L]$. The solution to this equation is $-4\phi = -2\beta L + 2n\pi$ or $\phi = \beta L/2 - n\pi/2$ which can be rewritten as

$$\tan\phi = \tan\left(\frac{\beta L}{2} - n\frac{\pi}{2}\right)$$

or

$$\frac{\alpha}{\beta} = \tan\left(\frac{\beta L}{2} - n\frac{\pi}{2}\right) \tag{6}$$

where *n* is an integer. Substitution of equation (1) and (2) into the equation (6) we obtain $\sqrt{(V-E)/V} = \tan(\beta L/2 - n\pi/2)$, which produce

$$E = V\cos^2\left(\frac{\beta L}{2} - n\frac{\pi}{2}\right) \tag{7}$$

Taking into account the definition of β in equation (2) and noticing that the electron energy here becomes discrete with quantum number n, then equation (7) can be rewritten as follows

$$E_n = V\cos^2\left(\frac{L}{2}\sqrt{\frac{2m}{\hbar^2}}E_n^{1/2} - n\frac{\pi}{2}\right) \tag{8a}$$

or

$$E_n^{1/2} = V^{1/2}\left|\cos\left(\frac{L}{2}\sqrt{\frac{2m}{\hbar^2}}E_n^{1/2} - n\frac{\pi}{2}\right)\right| \tag{8b}$$

It appears from the equation (8a) or (8b) that if $V \to \infty$ then the right hand approaches infinity, while the left hand side remains finite. To ensure both sides remain consistent, i.e., the right hand side stays finite the following condition must be fulfilled

$$\cos\left(\frac{L}{2}\sqrt{\frac{2m}{\hbar^2}}E_n^{1/2} - n\frac{\pi}{2}\right) \to 0$$

that causes $(L/2)\sqrt{2m/\hbar^2}\,E_n^{1/2} - n\pi/2 = \pi/2$ or

$$E_n = (n+1)^2\frac{\pi^2\hbar^2}{2mL^2} \tag{9}$$

Equation (9) is a standard solution for electron energies in a potential well having infinite wall height [1]. Thus, we can conclude that equation (8a) or (8b) is a general solution for electron energies in a potential well or arbitrary height and satisfying V > E.

Let us write the equation (8b) as follows

$$\frac{x}{V^{1/2}} = \left|\cos\left(\gamma x - n\frac{\pi}{2}\right)\right| \tag{10}$$

where $x = E_n^{1/2}$ and $\gamma = (L/2)\sqrt{2m/\hbar^2}$. The parameter γ is the wave number that is proportional to the film thickness. We can determine the solution of the equation (10) by plotting the curves of $x/V^{1/2}$ and $|\cos(\gamma x - n\pi/2)|$ simultaneously. The intersection of these two curves is a solution to x. Figure 2 shows the curves of these two functions as a function of x. We plot two curves of $x/V^{1/2}$ at two different *V* and plot two curves of $|\cos(\gamma x - n\pi/2)|$ at two different *γ*. Wider $|\cos(\gamma x - n\pi/2)|$ curve has been found using smaller γ. Since γ ∝ *L* then the wider curve belongs to thinner film.

Let us examine what happens if the potential barrier height fluctuates. The straight line curve will change its gradient. As a result, the solution for *x* also fluctuates. Let's examine the differences in fluctuations in *x* due to potential fluctuations. Distance of the intersection points of two straight lines with curved curve that has been obtained using large *γ* (large *L*) is smaller than the distance of intersection point of two straight lines with curved curves obtained using small γ) (small *L*). From these results we can conclude that the same fluctuations in the potential barrier produces greater energy fluctuations on the thinner film. In other words, the fluctuations of energies in the film are greater if the film is thinner.

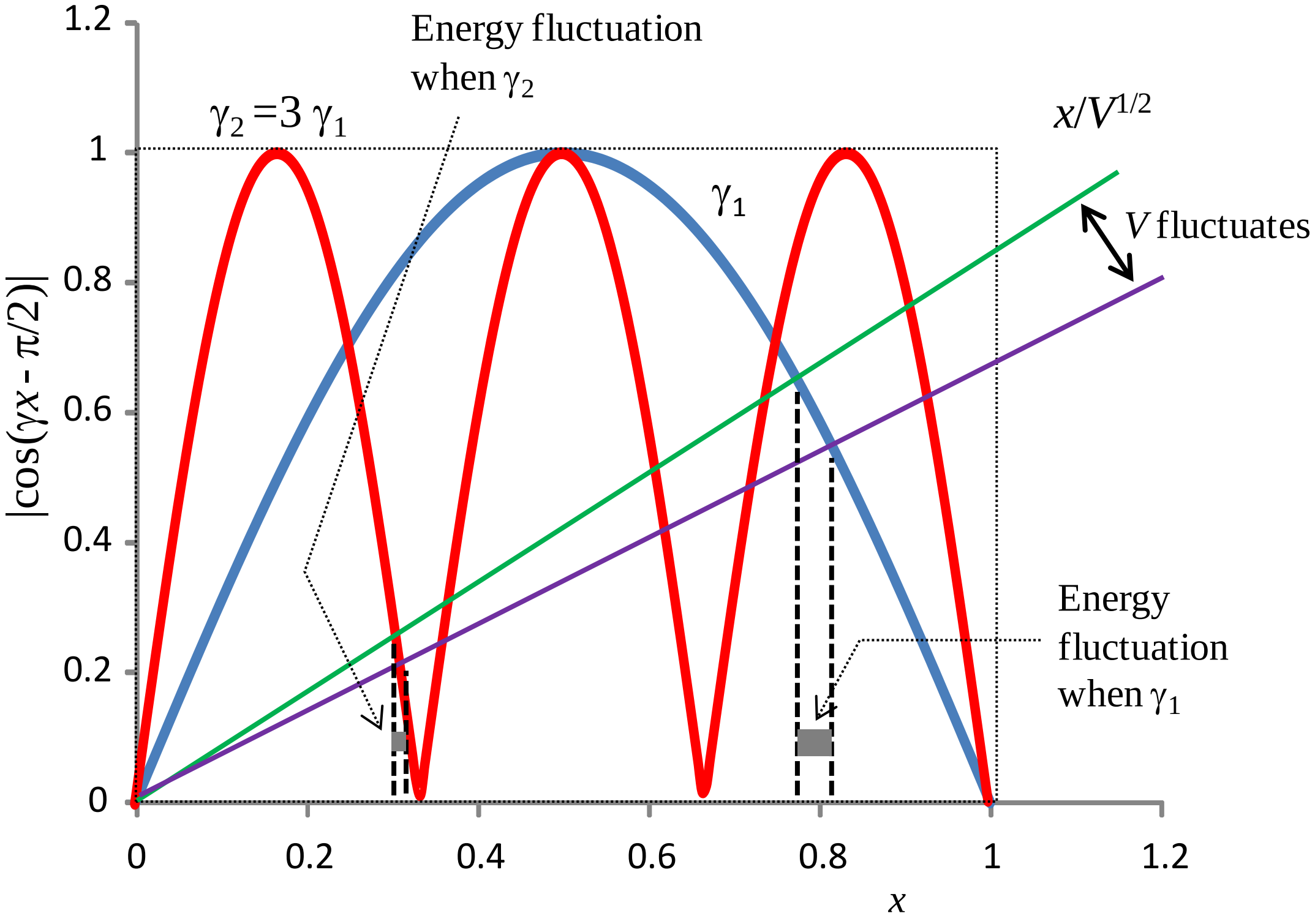

**Figure 2**. Curves of $x/V^{1/2}$ and $|\cos(\gamma x - n\pi/2)|$ as function of x. We plot two curves of $x/V^{1/2}$ using two different $V$. The intersection of the two curves is a solution to x.

Now we will look for more quantitative expression fluctuations. We start from the equation (8a). Suppose the fluctuation of potential barrier as much as δ*V* causing fluctuations of energy as $\delta E_n$. Thus we can write

$$E_n + \delta E_n = (V + \delta V)\cos^2\left(\gamma\sqrt{E_n + \delta E_n} - n\frac{\pi}{2}\right)$$

$$= V\cos^2\left(\gamma\sqrt{E_n + \delta E_n} - n\frac{\pi}{2}\right) + \delta V\cos^2\left(\gamma\sqrt{E_n + \delta E_n} - n\frac{\pi}{2}\right)$$

which gives the following equation

$$\delta E_n \approx \delta V\cos^2\left(\gamma\sqrt{E_n + \delta E_n} - n\frac{\pi}{2}\right)$$

$$\approx \delta V \cos^2\left(\gamma E_n^{1/2} - n\frac{\pi}{2} + \gamma\frac{\delta E_n}{E_n^{1/2}}\right) \tag{11}$$

where we already assume that $|\delta E_n/E_n^{1/2}| << 1$. We factorize the right hand side of equation (11) with the following standard trigonometry rules

$$\delta E_n = \delta V\left[\cos\left(\gamma E_n^{1/2} - n\frac{\pi}{2}\right)\cos\left(\gamma\frac{\delta E_n}{E_n^{1/2}}\right) - \sin\left(\gamma E_n^{1/2} - n\frac{\pi}{2}\right)\sin\left(\gamma\frac{\delta E_n}{E_n^{1/2}}\right)\right]^2$$

$$\approx \delta V\left[\cos\left(\gamma E_n^{1/2} - n\frac{\pi}{2}\right) - \gamma\frac{\delta E_n}{E_n^{1/2}}\sin\left(\gamma E_n^{1/2} - n\frac{\pi}{2}\right)\right]^2$$

$$\approx \delta V\left[\cos^2\left(\gamma E_n^{1/2} - n\frac{\pi}{2}\right) - 2\gamma\frac{\delta E_n}{E_n^{1/2}}\cos\left(\gamma E_n^{1/2} - n\frac{\pi}{2}\right)\sin\left(\gamma E_n^{1/2} - n\frac{\pi}{2}\right)\right]$$

$$\approx \delta V \cos^2\left(\gamma E_n^{1/2} - n\frac{\pi}{2}\right) - \gamma\frac{\delta V \delta E_n}{E_n^{1/2}}\sin 2\left(\gamma E_n^{1/2} - n\frac{\pi}{2}\right) \tag{12}$$

If rearranged the equation (14) yields

$$\delta E_n = \frac{\delta V \cos^2\left(\gamma E_n^{1/2} - n\frac{\pi}{2}\right)}{1 + L\sqrt{\frac{2m}{\hbar^2}}\frac{\delta V}{E_n^{1/2}}\sin 2\left(\gamma E_n^{1/2} - n\frac{\pi}{2}\right)} \tag{13}$$

Based on the equation (8a) we can write $\cos^2(\gamma E_n^{1/2} - n\pi/2) = E_n/V$, and $\sin 2(\gamma E_n^{1/2} - n\pi/2) = 2\sin(\gamma E_n^{1/2} - n\pi/2)\cos(\gamma E_n^{1/2} - n\pi/2)$ = $2\sqrt{E_n/V}\sqrt{1 - E_n/V}$ so that equation (13) can be rewritten as

$$\delta E_n = \frac{\delta V \frac{E_n}{V}}{1 + L\sqrt{\frac{2m}{\hbar^2}}\frac{\delta V}{E_n^{1/2}} \times 2\sqrt{\frac{E_n}{V}}\sqrt{1 - \frac{E_n}{V}}}$$

$$= \frac{\frac{\delta V}{V}}{1 + 2L\sqrt{\frac{2m}{\hbar^2}}\frac{\delta V}{V}\sqrt{V - E_n}} E_n$$

Furthermore, we approximate En in the equation above with $E_n$ for infinite wall height as appeared in equation (9) to obtain more explicit forms of energy fluctuations as follows

$$\delta E_n \approx \frac{(n+1)^2 \pi^2 \hbar^2 / 2m}{1 + \left[ \frac{2\sqrt{2m(V - E_n)}}{V\hbar} \delta V \right] L} \left( \frac{\delta V}{V} \right) \left( \frac{1}{L^2} \right) \tag{14}$$

It appears from the equation (14) that the greater the energy fluctuations if the film thinner. If we assume the walls potential height is high enough so that the discrete energies in the well are quite small compared potential height, or $E_n << V$ and potential fluctuations on the walls is very small compared with the $\sqrt{V}$ then we can make an approximation

$$\delta E_n \approx \left[ \frac{(n+1)^2 \pi^2 \hbar^2 (\delta V / V)}{2m} \right] \frac{1}{L^2} \tag{15}$$

It is clear that fluctuations in energies change according to the inverse square of the film thickness. Also shown in equation (15), the fluctuations in the energies are proportional to relative fluctuation in the wall potential. The greater this ratio, the smaller the energy fluctuates.

## 2.2 Effects on Lattice Constant

Fluctuations in the energy states are manifestation of the fluctuations in the electron wave functions in the well. The electron density in the wells is $N|\psi|^2$, $N$ is the concentration of electrons. Accordingly, fluctuations in the electron density in the well is

$$\delta\left(N|\psi|^2\right) = N\sum_n \frac{\partial|\psi|^2}{\partial E_n}\delta E_n = \left[N\sum_n f(n)\frac{\partial|\psi|^2}{\partial E_n}\right]\frac{1}{L^2} \tag{16}$$

with f(n) is a function of the index n. The root mean square fluctuations of the electron density is

$$\sqrt{\left\langle\left[\delta\left(N|\psi|^2\right)\right]^2\right\rangle} = \sqrt{\left\langle\left[N\sum_n f(n)\frac{\partial|\psi|^2}{\partial E_n}\right]^2\right\rangle}\frac{1}{L^2}$$

If $\rho_0$ is the average velocity of electrons in the wells, the electron density fluctuations can be written as

$$\rho = \rho_0 \pm \sqrt{\left\langle\left[N\sum_n f(n)\frac{\partial|\psi|^2}{\partial E_n}\right]^2\right\rangle}\frac{1}{L^2} \tag{17}$$

The ions that compose material are unified by attraction force between the positively charged ions and the electron.  The magnitude of attractive force is roughly proportional to the electron density. In the presence of fluctuations in the electron density, then the force between the ions and electrons also fluctuates. Changes in force experienced by ion is

$$\Delta F = \frac{\rho - \rho_0}{\rho_0}F_0 = \mp\frac{F_0}{\rho_0}\sqrt{\left\langle\left[N\sum_n f(n)\frac{\partial|\psi|^2}{\partial E_n}\right]^2\right\rangle}\frac{1}{L^2} \tag{18}$$

with$F_0$ is the force when the fluctuation absent. This force produces a lattice constant changes that can be predicted by the equation of Young's modulus ($Y \approx (F/a^2)/(\Delta a/a)$). It appears that the lattice constant change is proportion to changes in force. Thus, the lattice constant change satisfies $\delta a = -\mu / L^2$, where $\mu$ is a parameter that may be different for different materials. Because fluctuations can increase or decrease the potential, the parameter $\mu$ can be positive or negative number. Thus the fluctuations of electron states can cause the distance between atoms expand or shrink.

Equation (14) or (15) has been derived for thin films. The energy states of electrons in a thin film are inversely proportional to the square of the film thickness. The resulting change in the lattice constant is inversely proportional to the square of the film thickness. For particle (sphere), the energy states are inversely proportional to the diameter of particles [7]. With these properties, it is expected that the lattice constant changes in the particle is inversely proportional to the square of the diameter, or $\delta a = -\mu / D^2$. As a result, we get equations describing the variation in the lattice constant of the thins or nanoparticles as

$$a_0 = a_0(\infty) - \frac{\mu}{L^2} \tag{19a}$$

$$a_0 = a_0(\infty) - \frac{\mu}{D^2} \tag{19b}$$

Leont'yev et al. have observed dependence of the lattice constant of platinum supported carbon with a particle size between 2-28 nm by X-ray diffraction. They obtained a fitting equation $a = a_0 + b/D$ with $a_0 = 3.9230 \pm 0.0017$ A and $b = -0.0555 \pm 0.0067$ $\mathrm{nm}^{-1}$($D$ in nm) [3].

## 3. Confirmation with Experimental Data

Symbols in Figure 3 are the measurement results of lattice constant in platinum nanoparticles on the (222) and (422) orientations as reported by Soliard and Flueli [2]. The presented data are the measurement results at 300 K. The curve in the figure is obtained using equation (19b). The fitting equations which give the smallest error are ($a_0 = 3.926 - 0.0473/D^2$ angstrom, $\sigma = 0.00142$) and ($a_0 = 3.917 - 0.0568/D^2$ angstrom, $\sigma = 0.00119$) for (222) and (422) orientations, respectively, and $\sigma$ is the standard deviation. Soliard and Flueli have reported that the lattice constant depends on the inverse of diameter [2]. For comparison we also determine the fitting curve that varies with inverse of diameter as proposed by Soliard and Fueli. The fitting equations that give the smallest error are ($a_0 = 3.9323 - 0.0364/D$ angstrom, $\sigma = 0.00104$) and ($a_0 = 3.9247 - 0.0426/D$ angstrom, $\sigma = 0.00078$) for (222) and (422) orientations, respectively. From this result it appears that fitting with the function which is inversely proportional to the radius or inversely proportional to the square of the radius can be accepted because they both give very small standard deviation and almost equal.

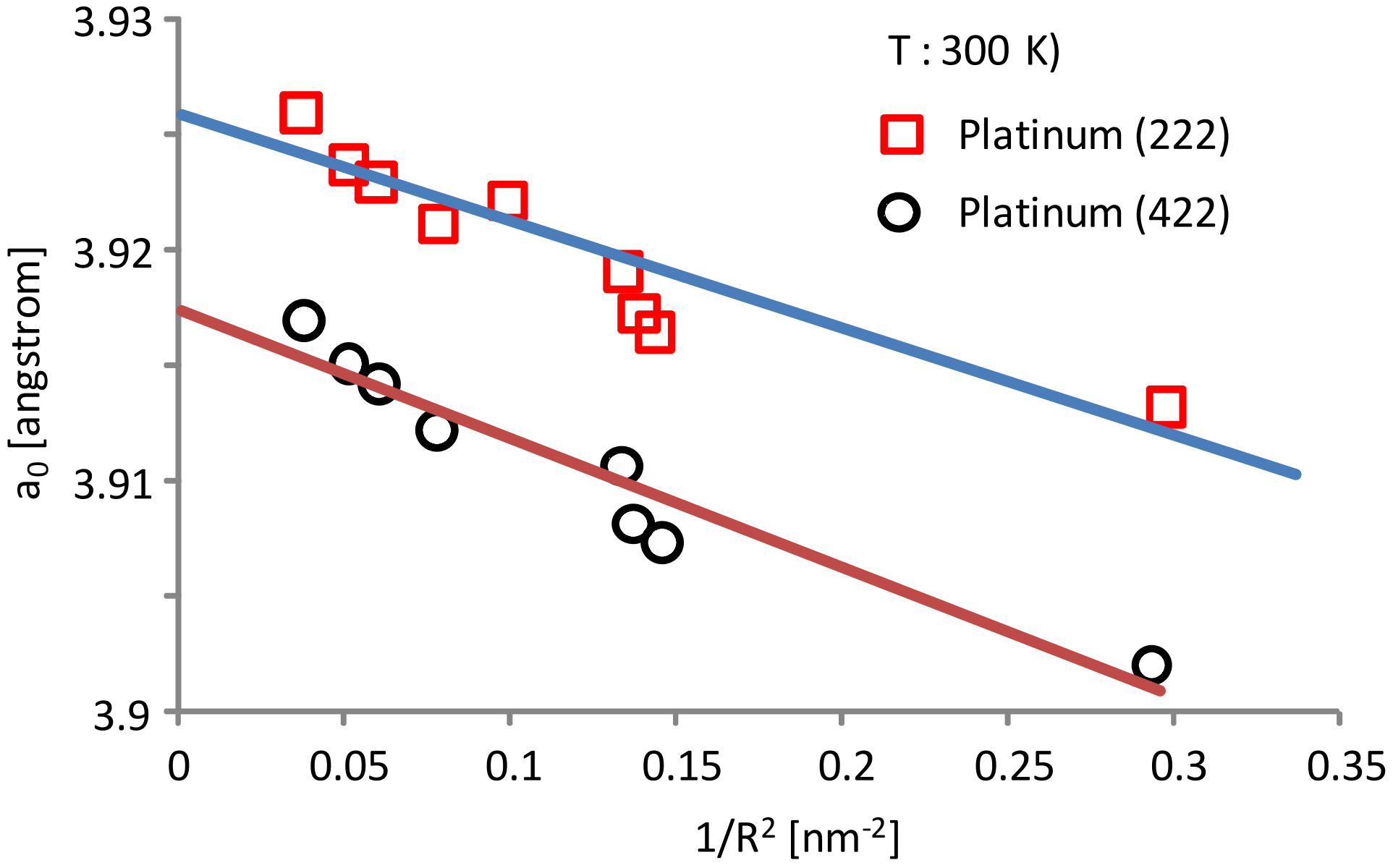

**Figure 3.** Symbols are lattice constants platinum nanoparticles at a temperature of 300 K and in the (222) and (422) orientations as reported by Soliard and Flueli [2]. The curve has been obtained by using equation (19b).

Symbols in Figure 4 are the lattice constants of gold nanoparticles on the (222) and (422) orientations as reported by Soliard and Flueli [2]. The presented data are the measurement results at 300 K. The curve has been obtained by using equation (19b). The fitting equations that give the smallest error are ($a_0$ =4.0717 – 0.0714/$D^2$ angstrom, $\sigma$ = 0.00207) and ($a_0$ =4.0682 – 0.0673/$D^2$ angstrom, $\sigma$ = 0.00284) for (222) and (422) orientations, respectively. Soliard and Flueli have reported that the lattice constant depends on the inverse of diameter [2]. For comparison we also determine the fitting curve that varies with inverse of diameter as proposed by Soliard and Flueli [2]. The fitting equations that give the smallest error are ($a_0$ =4.0792 –

0.05126/*D* angstrom, σ = 0.00192) and ($a_0$ =4.0755 – 0.04902/*D* angstrom, σ = 0.00226) for (222) and (422) orientations, respectively. From this result it also appears that fitting with the function which is inversely proportional to the radius or inversely proportional to the square of the radius can be accepted.

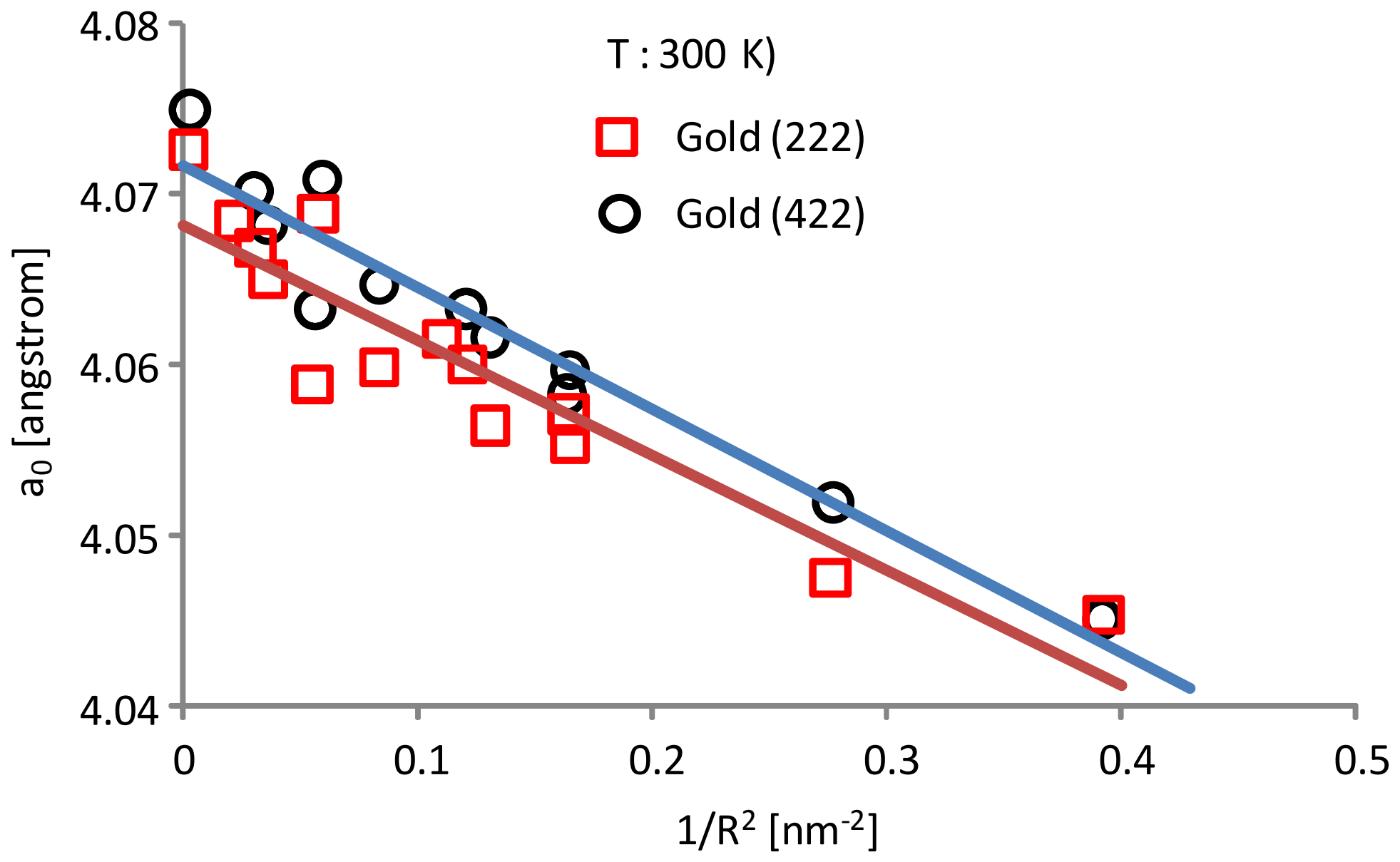

**Figure 4.** Symbols are lattice constants gold nanoparticles at a temperature of 300 K and in the (222) and (422) orientations as reported by Soliard and Flueli [2]. The curve has been obtained by using equation (19b).

Symbols in Figure 5 ate the result of the measurement of lattice parameters for the carbon-supported platinum nanoparticles as reported by Leontyev et al [3]. The fitting equation that gives the smallest error is ($a_0$ =3.9173 – 0.1170/$D^2$ angstrom, σ = 0.0040). Leont'yev et al have that the lattice constant depends on the inverse of diameter. We also determine the fitting curve that varies with inverse of diameter as proposed by Leontyev et al [3]. The fitting equation that give the smallest error ($a_0$ =3.9247 – 0.06866/*D* angstrom, σ = 0.0212). It appears that the results of fitting a function that varies with inverse of square diameter as given by equation (19b) gives a smaller standard deviation. In other words, the result of fitting using the equation (19b) is more accurate than the result of fitting using equation that varies with inverse of diameter.

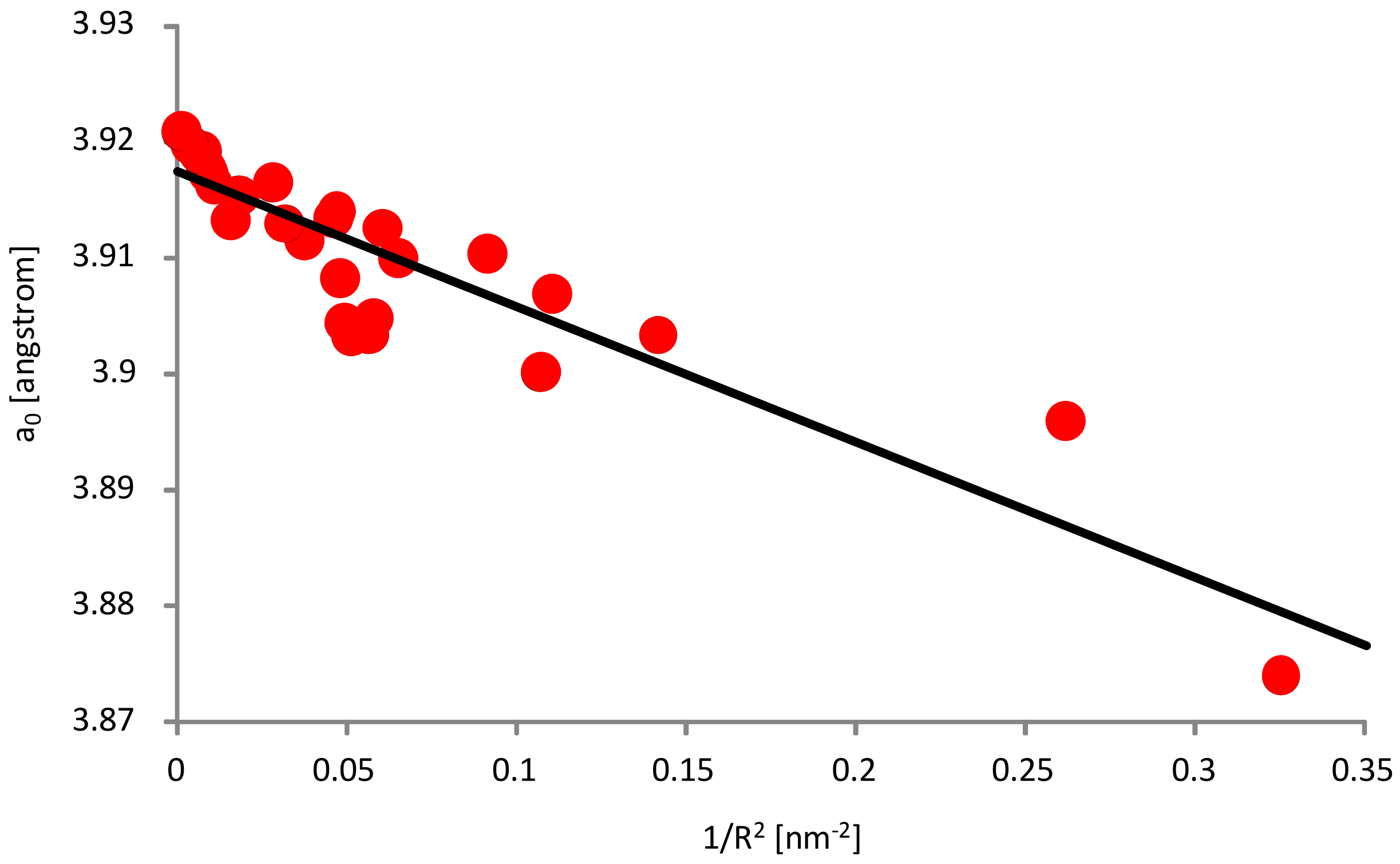

**Figure 5.** Symbols are lattice constants carbon-supported platinum nanoparticles as reported by Leontyev et al [2]. The curve has been obtained by using equation (19b).

Symbols in Figure 6 are the result of the measurement of lattice parameters for ceria nanoparticles as reported by Chen et al [4]. Square symbols are data for nanocrystal synthesized by the method of micelle template, while the circle symbols are data for nanoparticles synthesized with a simple precipitation method. The fitting curve that gives the smallest error for micelle template particles is ($a_0$ =5.4168+0.1066/$D^2$ angstrom, $\sigma$ = 0.00071) and for simple precipitated particles is ($a_0$ =5.4085-0.2284/$D^2$ angstrom, $\sigma$ = 0.00146). Also seen here that the standard deviations obtained for both fittings are very small, which proves that the proposed model is quite good in explaining a number of experimental data.

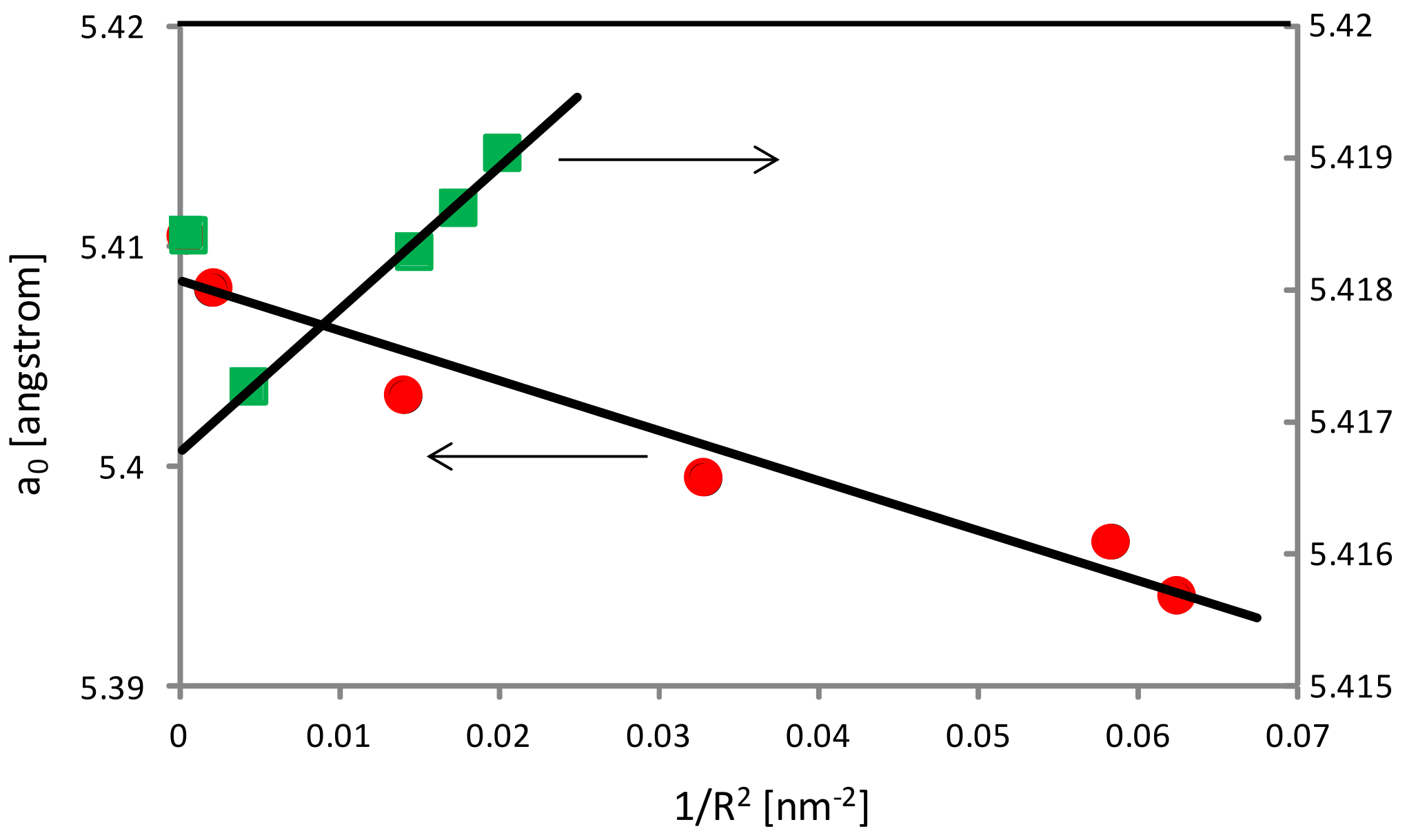

**Figure 6.** Symbols are the result of the measurement of lattice parameters for ceria nanoparticles as reported by Chen et al [4]. Square symbols are data for nanocrystal synthesized by the method of micelle template, while the circle symbols are data for nanoparticles synthesized with a simple precipitation method. The curves have been obtained by using equation (19b).

## 4. Conclusion

We have successfully developed a model for predicting fluctuations eigen states of electrons in thin films. We have also succeeded in establishing the equations that describe the effect of film thickness or particles diameter on a lattice constant variation. The obtained equation can explain well several experimental data on the variations of lattice constant when particle changes.

## Acknowledgement

This work was supported by a research grant (No. 310y/I1.C01/PL/2015) from the Ministry of Research and Higher Education, Republic of Indonesia, 2015-2017.

## References

[1] S. Gasiorowicz, *Quantum Physics*, 3rd Edition, New York: John Wiley (2003)

[2] S. Soliard and M. Flueli, *Surface Science* **156**, 487 (1985).

[3] I.N. Leontyev, A.B. Kuriganova, N.G. Leontyev, L. Hennet, A. Rakhmatullin, N.V. Smirnova, and V. Dmitriev, *RSV Advances* **4**, 35959 (2014).

[4] L. Chen, P. Fleming, V. Morris, J. D. Holmes, and M. A. Morris, *J. Phys. Chem. C* **114**, 12909 (2010).

[5] T. P. Sarkar, K. Gopinadhan, M. Motapothula, S. Saha, Z. Huang, S. Dhar, A. Patra, W. M. Lu, F. Telesio, I. Pallecchi, Ariando, D. Marré & T. Venkatesan, *Scientific Reports* **5**, 13011 (2015).

[6] H. Zhang, B. Chen, and J.F. Benfield, *Phys. Chem. Chem. Phys*. **11**, 2553 (2009).

[7] M. Abdullah (ed), *Nanosains & Nanoteknologi*, Bandung: Penerbit ITB (2015).